\documentclass[a4paper,12pt]{article}

\usepackage{graphicx} \usepackage{amsmath}

\title{Spreading of Fluids on Solids Under Pressure: Effect of Slip}
\author{Soma Nag$^1$, Tapati Dutta$^2$ and Sujata Tarafdar$^{1*}$}

\begin{document}

\maketitle
\noindent
$^1$Condensed Matter Physics Research Centre, Physics Department\\ Jadavpur University, Kolkata 700032, India\\ 
$^2$Physics Department, St. Xavier's College, Kolkata 700016, India\\
\noindent
${^*}$ Corresponding author: Email: sujata$\_$tarafdar@hotmail.com,\\  Phone:+913324146666(extn. 2760)

\begin{abstract} \noindent
 Spreading of different types of fluid on substrates under an impressed force is an interesting problem. Here we study spreading of  four fluids, having different hydrophilicity and viscosity on two substrates - glass and perspex, under an external force. The area of contact of fluid and solid is video-photographed and its increase with time is measured. The results for different external forces can be scaled onto a common curve. We try to explain the nature of this curve on the basis of existing theoretical treatment where either the no-slip condition is used or slip between fluid and substrate is introduced. We find that of the eight cases under study, in five cases quantitative agreement is obtained using a slip coefficient. \\
\end{abstract}
\noindent {\bf Keywords}: Spreading, squeeze film, viscous flow, slip\\
\noindent {\bf PACS Nos.}: Wetting in liquid-solid interfaces - 68.08.Bc\\
                         : Viscosity, experimental study - 66.20.Ej

\section{Introduction}
Wetting and spreading\cite{degen, zisman, bonn} are very common everyday phenomena, but there are still a lot of open problems in this field. A simple example is - what happens when a fluid is forced to spread under an externally impressed force ? There is no fully satisfactory answer to be found in the otherwise comprehensive literature on the topic, though such experiments are routinely used to characterize viscoelastic materials.

In the present paper, we report a set of very simple experiments on the so-called 'squeeze film' and try to find a theoretical framework to explain our results. The problem is briefly described as follows - a small droplet of a liquid, placed on a solid plate will take the shape of a section of a sphere. Under equilibrium, the well known Young-Laplace relation holds, involving an equilibrium angle of contact $\theta_e$.The dynamic condition  while the fluid is spreading, i.e. the area of contact of the fluid and solid is increasing in time, is also well studied. There are a number of excellent reviews summarising studies on - the rate of spreading, the forces responsible and other interesting features \cite{degen, bonn, engman, asthana}. We now look at a simple extension of the problem - what happens if the fluid is forced to spread beyond its equilibrium area, under a constant weight placed on top of it. A preliminary report was published by our group in \cite{apsurf}.
In this paper we study four different fluids on two different solids, so as to have different combinations of viscosities of fluids as well as different degrees of  hydrophilicity of the  fluids and solids. The results are analyzed on the basis of viscous energy dissipation with and without slip\cite{tabor,engman}.

We find that while the no-slip result gives a fair quantitative estimate of the contact area as function of time for only one case, it is possible to get a better agreement in 5 of the 8 cases, on introducing a slip coefficient. Of course, this involves introduction of a free parameter (the slip coefficient) whereas the no-slip analysis did not. The slip coefficient appears to depend more on the solid surface than the fluid, and is an order of magnitude lower for the perspex plate. The results for spreading of the same fluid on different surfaces is quite different. There are 3 cases where the agreement between theory and experiment is unsatisfactory and cannot be improved by introducing slip, so other effects are also at play. We describe the experimental setup followed by the theory and present detailed results in the third section. Finally we compare experimental  results with theory and discuss the implications.

\section{Experimental}

The experimental fluids are castor oil, linseed oil, glycerol and ethylene glycol. The choice of fluids is motivated by the following idea. While castor oil and linseed oil are insoluble in water, with a low    dielectric constant $\epsilon$ the other two fluids ethylene glycol and glycerol having high dielectric constants, are soluble in water. ($\epsilon$ = 4.67, 3.35, 47.70 and 43.0 respectively for castor oil,linseed oil, glycol and glycerol). Again, one fluid belonging to each type according to hydrophilicity, has high viscosity($\mu$), namely castor oil and glycerol($\mu$ for castor oil=986-451 cp for temp 20-30c,glycerol is 1490-629 cp for temp 20-30c), while the others have much lower viscosity($\mu$ for glycol is 19.9-9.13 cp for temp 20-40c , linseed oil=33.1-17.6 cp for temp 20-50c). 

\begin{figure}[ht]
\begin{center}
\includegraphics[width=14.0cm, angle=0]{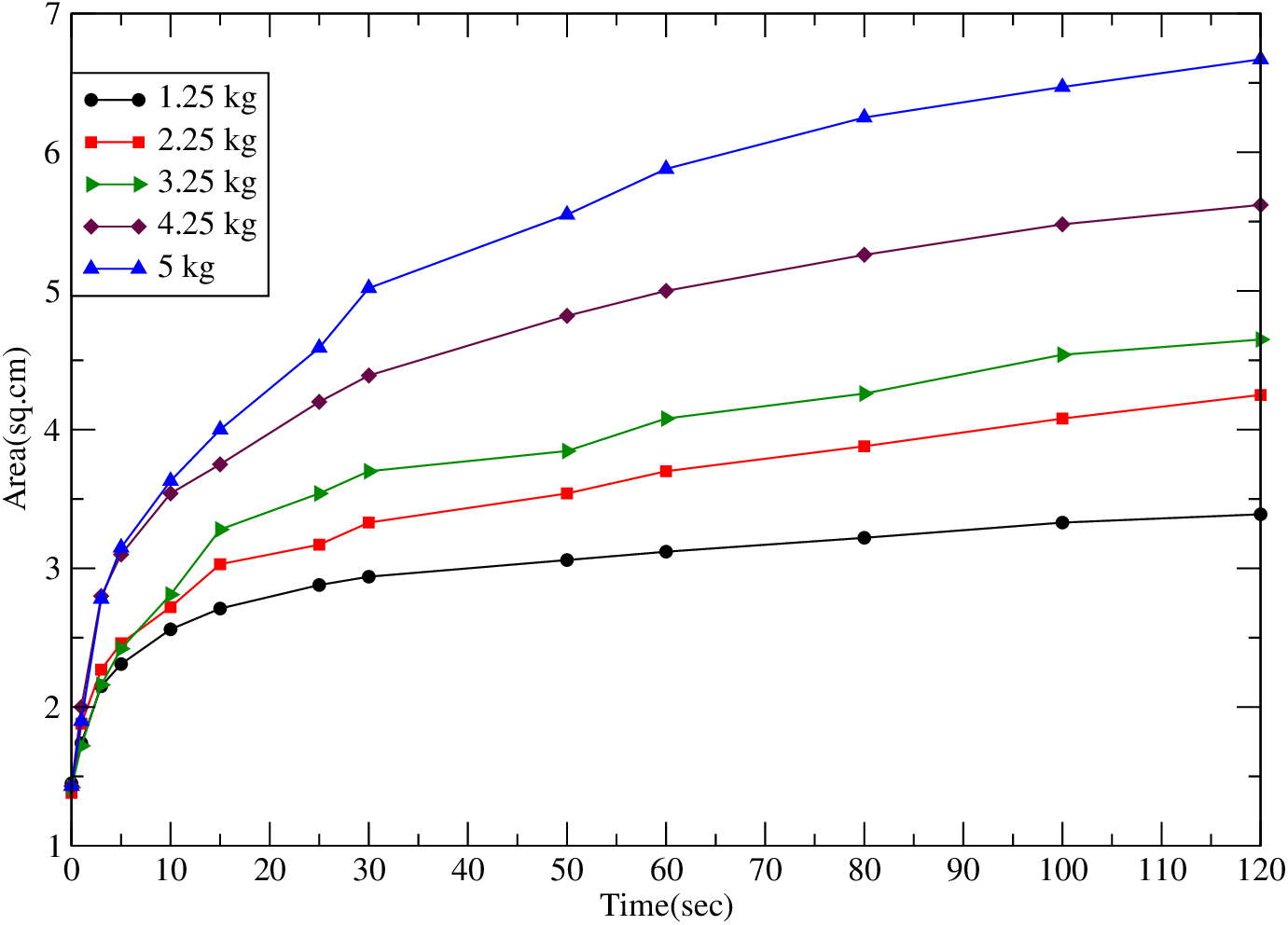}
\end{center} \caption{This figure shows area vs. time graph for Castor oil on perspex.} \label{tacp}
\end{figure}

\begin{figure}[ht]
\begin{center}
\includegraphics[width=14.0cm, angle=0]{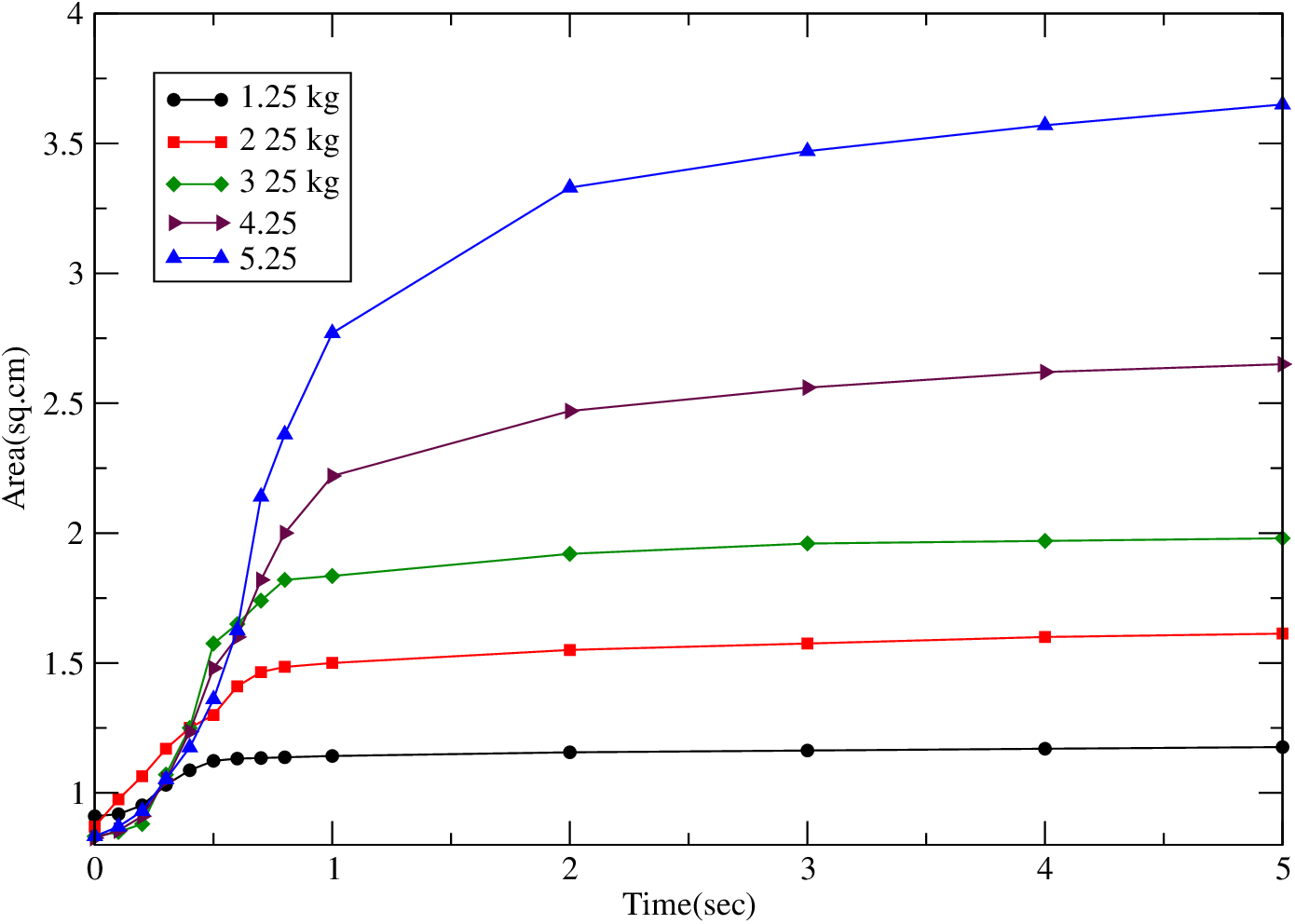}
\end{center} \caption{This figure shows area vs.time graph for glycol on perspex.} \label{tagp}
\end{figure}

A constant volume $V_f$ of the fluid is measured with a micropipette and placed on the lower plate. The upper plate is then
placed on the drop and a load of M kg. (M = 1 to 5) is placed on it. The pair of plates are $\sim$ 1 cm thick, one pair is made of float glass and the other of perspex. A video camera, placed below the lower glass plate records the slow increase in area. The video clips are analyzed using Image Pro-plus software.  We find that the area of the drop in contact with the plate increases with time at first. Then after some time it reaches  a saturation value and does not increase significantly with time any more. The time of saturation and area of saturation are different for different fluid-solid combinations. We plotted the area vs. time graph for all fluid-solid combinations.  Fig(\ref{tacp}) and fig(\ref{tagp}) show respectively area vs. time graphs for Castor oil and glycol respectively on perspex for loads from 1-5 kg. Interestingly when the data for area are divided by the load mass M, the curves come very close to each other, almost collapsing to a single master curve. Again, for castor 
oil on perspex area increases for a long time. It takes several minutes to saturate approximately and continue to increase very slowly even after 5 minutes.For glycol on both the substrates, on the other hand, the area increases very fast and saturates in a few seconds. Video photos are taken initially, at the rate of 10 frames/sec, in this case.

\section{Theory}

We discuss in this section standard formulations for this kind of problem.
The 'squeeze flow' usually discussed in chemical engineering literature is  different from our experiment, ours is a constant volume squeeze, where whole mass of fluid taken is always confined within the two plates. In the usual experiment a {\it constant area} setup is used, where both plates are immersed in the fluid, the distance between the plates as function of time is measured as a force is applied on the upper plate.

First we outline the steps worked out for the no-slip case, i.e. assuming that the layer of fluid next to the static plate has zero velocity. The analysis roughly follows \cite{tabor}.

\subsection{No-slip}
We assume the fluid to be Newtonian and incompressible with constant volume $v_f$. Ours is a constant force set up, the force $Mg$ acts on the fluid layer. The velocity first builds up, reaches a maximum and then starts to reduce  as it squeezes the liquid film. The details are presented in \cite{apsurf}, here we consider the velocity decreasing regime only. We equate the loss of potential energy of the loaded plate with the work done against the viscous resistance. With $h$ and $r$ representing the instantaneous thickness and radius of the film respectively, $p$ the pressure and $p_A$ the atmospheric pressure. In moving over a distance dh the work done is given by

\begin{equation}
    dh  \int (p-p_A)2\pi rdr
\end{equation}

The loss of potential energy is

\begin{equation}
    Mg(h-(h-dh))=Mgdh
\end{equation}

Now the pressure p at any point r from the center of the upper plate for any given distance h, where the velocity of the upper plate $V=-dh/dt$ can be calculated as 

\begin{equation}
  p-A = {\frac{3\mu V(R^2-r^2)}{h^3}}
\end{equation}

Therefore, we can write,
 
\begin{equation}
  \int Mg dh = \int dh  \int  (p-p_A)   2  \pi  r  dr
\end{equation}

Substituting for $(p - p_A)$, we  get,

\begin{equation}
   Mg = \int { \frac {3 \mu V (R ^ 2 - r ^ 2 )} { h ^ 3 } }    2  \pi  r  dr       
\end{equation}

Which , on integration gives,

\begin{equation}
    Mg = -{\frac{3\mu v_f^2}{2\pi h^5}}{\frac{dh}{dt}}
\end{equation}

\begin{equation}
   \int_0^t  Mg dt  = - { \frac{3\mu v_f^2}{2\pi}}  \int_1^2 \frac{dh}{h^5}
\end{equation}

\begin{equation}
    Mgt  =  {\frac{ 3 \mu(v_f)^2 }{ 8\pi }}   [{ \frac{1}{h_2^4} }- { \frac{1}{h_1^4} }]
\end{equation}

\begin{equation}
    Mgt  =  {\frac{ 3 \mu}{ 8\pi v_f^2 }}   ( A_2^4 - A_1^4 )
\end{equation}

The right hand expression  depends on the impressed force $Mg$ through the area of contact $A(t)$. 
If we divide both sides by $M$, the LHS is independent of $M$. The RHS is now a function of $A(M),M$ and $t$. We call this function $G(M,t)$.
We calculate the RHS from experimental data and compare with the LHS. If the principles on which the derivation is based are correct, the RHS calculated from experimental results should be independent of $M$. That is the curves $G(M,t)$ vs. 
$t$ should collapse to a single curve and be numerically equal $gt$.

so, we test the result

\begin{equation}
    gt  =   G(M,t)
 \label{gmt} \end{equation}

by plotting the LHS and RHS separately on the same graph and check how well they agree. Of course the agreement can be expected only for the time interval before the area of contact saturates, since the LHS goes on increasing linearly.

\subsection{Analysis of Experimental Results assuming no-slip}
\begin{figure}[ht]
\begin{center}
\includegraphics[width=14.0cm, angle=0]{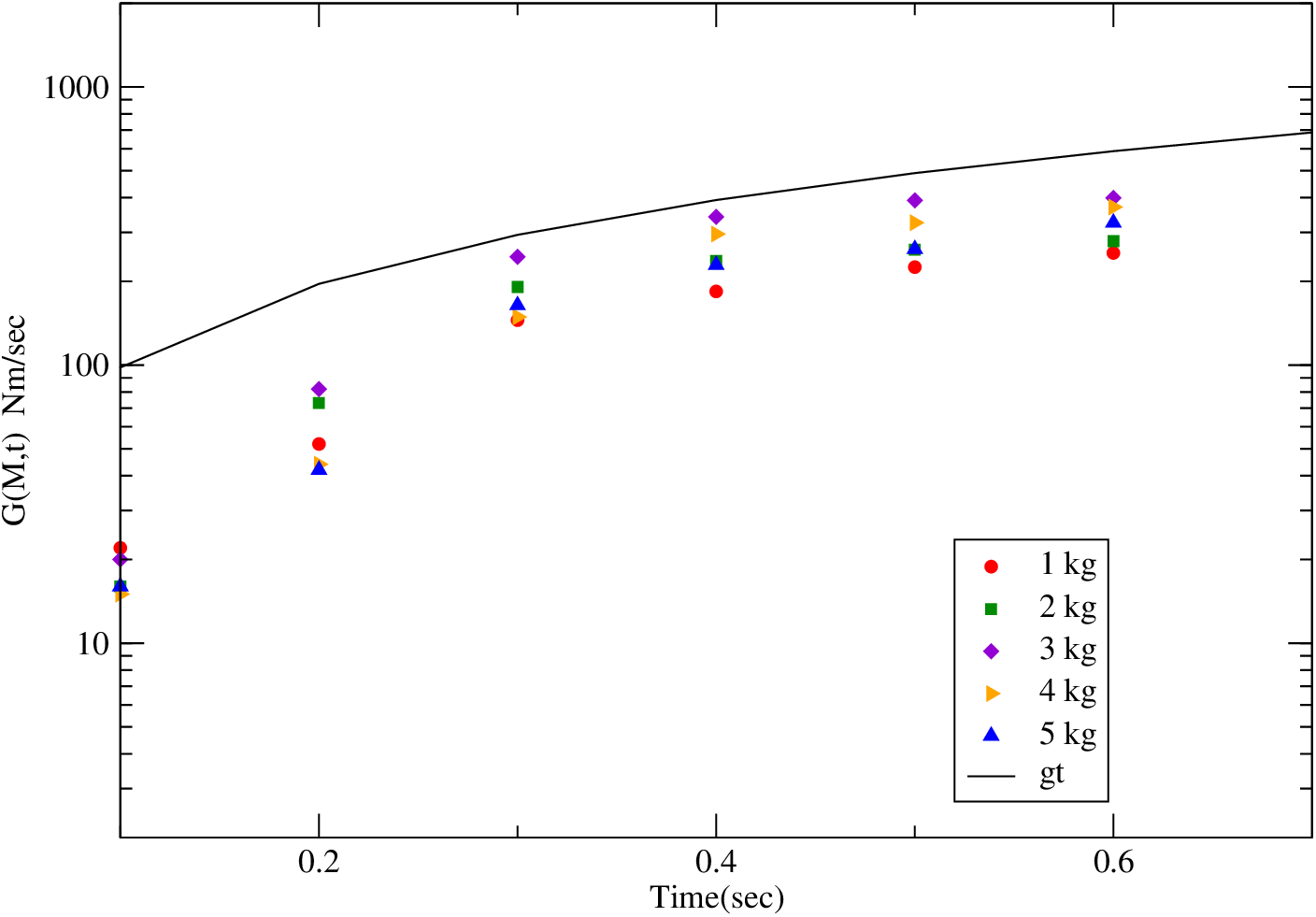}
\end{center} \caption{This figure tests gt=G(M,t) results for glycerol on glass} \label{gmt_glcrg}
\end{figure}

The results for glycerol on glass are shown in fig(\ref{gmt_glcrg}).
We see that for this combination assuming no slip gives an  acceptable quantitative agreement between the lhs and rhs of eq(\ref{gmt}).  This calculation involves no adjustable parameter. It may further be noted that the data for different loads collapse almost to a single line, on scaling by $M$.
However, for the other fluid-solid combinations, such quantitative agreement is not found. We try to see now if this discrepancy may be attributed to the effect of slip at the fluid and solid interface.

 We follow the equation given by \cite{stefan} listed in \cite{engman} for this situation.

\subsection{Effect of slip}

Relative motion between a solid surface and material contacting it (wall slip) is a phenomenon observed with many materials \cite{engman}. In
the last section we assumed that the layer of the liquid touching the lower fixed plate has zero velocity. So that the applied force is entirely carried away by viscous dissipation. But experiments show that there may be cases where the layer of the liquid having contact with the substrate may have finite velocities, which may affect the spreading of the liquid drop. In these cases, shear stresses at the sample-plate interfaces may vanish (for perfect slip) or reduce (for partial slip). The amount of slip depends not only on the liquid tested, but also on the material and roughness of the substrates. We introduce partial slip in the analysis as follows.

\subsection{Partial slip}

No slip and perfect slip are two limiting cases of slip phenomenon. To explain our results, we take partial slip into consideration. Stefan\cite{stefan} gave a slip law of the form ,
 
\begin{equation}
      v =  \beta  *  \tau
\end{equation}

where,  $\beta$ is the slip coefficient and $\tau$ is the wall shear stress . The value of $\beta$ is zero for no slip and is infinity for perfect slip cases. It may have any value in between, for partial slip cases, depending on the liquid and the substrate.

With the introduction of $\beta$,the force equation becomes,  

\begin{equation}
 Mg = \int \frac{3 \mu v_f (R^2 - r^2) 2 \pi r dr}{h^3(1 + {6\beta \mu}/h) }
\end{equation}

\begin{equation}
   \int Mgdt = -{\frac{3\mu v_f^2}{2\pi}} \int{\frac{dh}{h^5+6\beta \mu h^4}}
\end{equation}

\begin{equation}
   Mgt  =  -{\frac{3\mu v_f^2}{2\pi}} \int{\frac{dh}{h^5+6\beta \mu h^4}}
\end{equation}

Dividing by $M$.

\begin{equation}
  gt  =  R(M,t)
  \label{rmt}
\end{equation}

where, the integral can be evaluated as
\begin{equation}
  R(M,t)=-{1/k^4}*[k^3/h^3+3k^2/2h^2-ln(1+k/h)]_{h1}^{h2}
\end{equation}

with $k = 6 \beta \mu$

The above force equation, on integration,  gives us the  function $R(M,t)$  similar to $G(M,t)$ for the no-slip case. We fit the slip coefficient $\beta$ so as to get the best fit between the left and right sides of equation \ref{rmt}. the results 
are shown in the next section.
\subsection{Results with partial slip} 

 Table(\ref{tabl}) shows values of $\beta$ for different fluid-solid combinations. We find that the values of $\beta$ ranges from $10^{-3}$ to $10^{-5}$ for glass and from $10^{-5}$ to $10^{-6}$ for perspex. So glass  has a larger $\beta$ than perspex. For a definite substrate the values of $\beta$ are  different for  different  liquids. In Table 1 we also show the initial contact area $A_0$ (before loading) and the viscosities and dielectric constants of the fluids and substrates.
                                                            
\begin{figure}[ht]
\begin{center}
\includegraphics[width=14.0cm, angle=0]{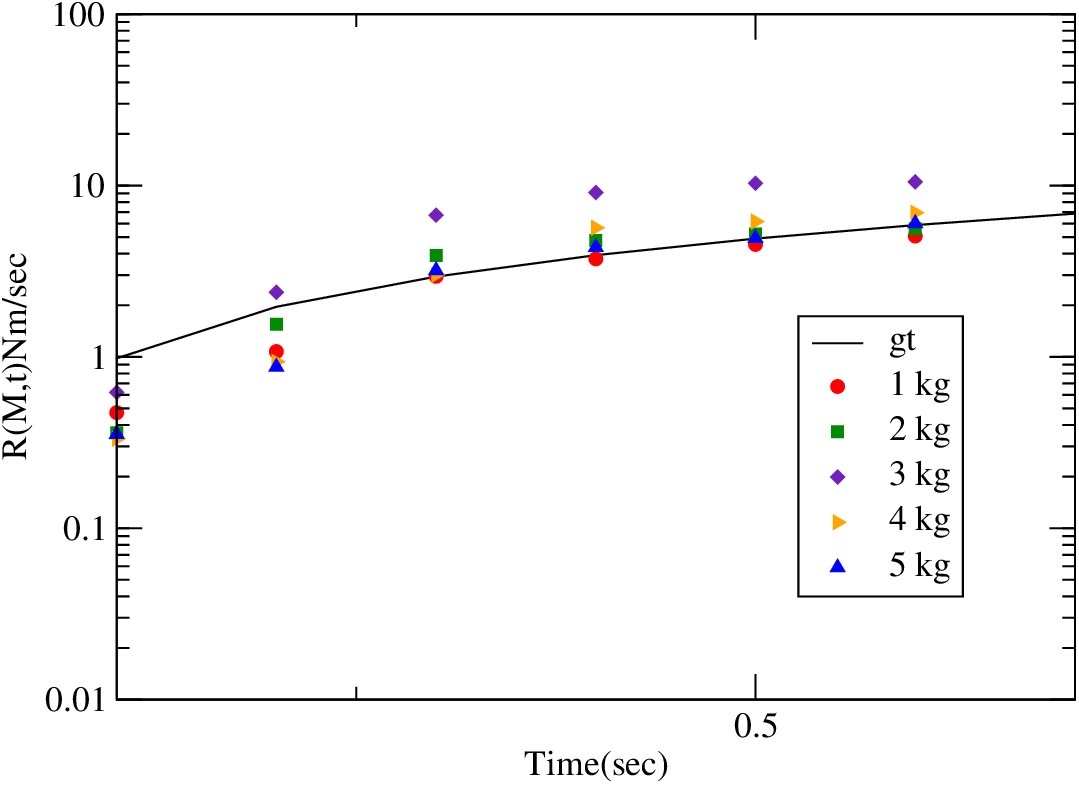}
\end{center} \caption{This figure tests gt = R(M,t) results for glycol on glass. Introduction of slip coefficient $\beta$ gives a good agreement.} \label{rmt_gg}
\end{figure}

\begin{figure}[ht]
\begin{center}
\includegraphics[width=14.0cm, angle=0]{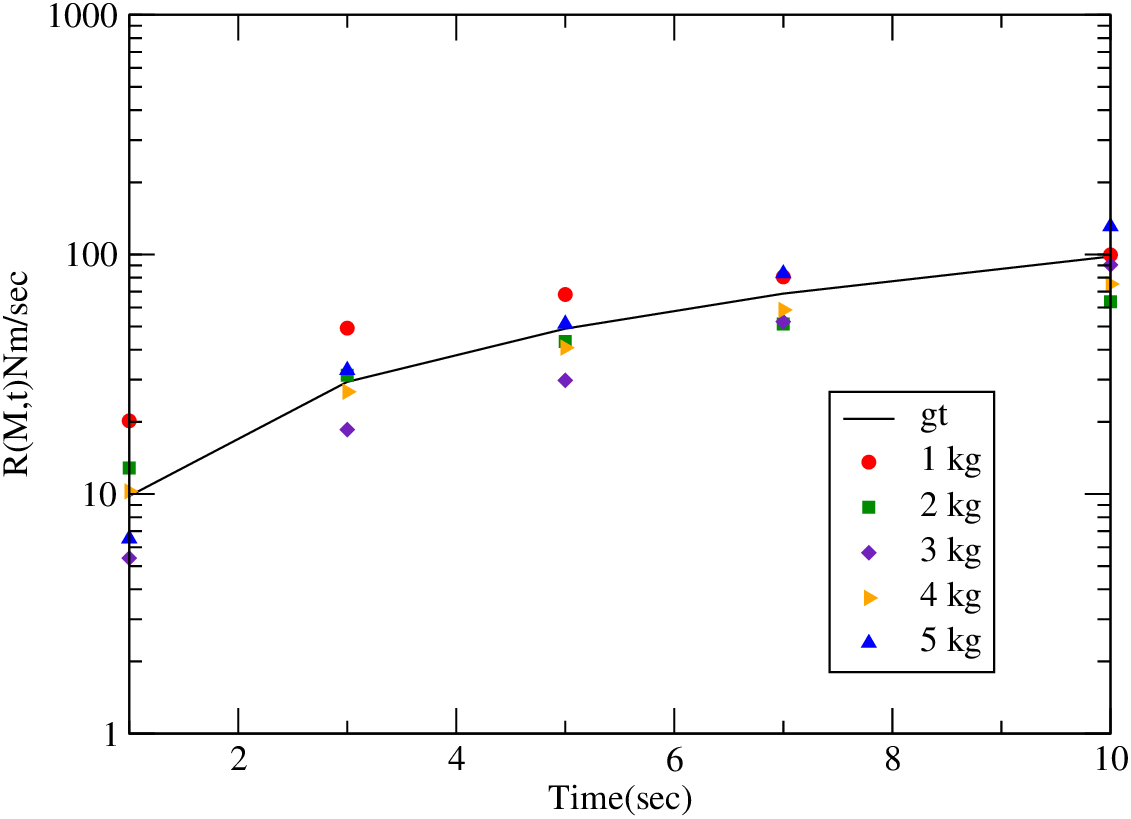}
\end{center} \caption{This figure tests gt = R(M,t) results for castor oil on perspex. Introduction of slip coefficient $\beta$ gives a good agreement.} \label{rmt_cp}
\end{figure}

Figures (\ref{rmt_gg}) and (\ref{rmt_cp}) show respectively the graphs comparing $R(M,t)$ and $gt$ for ethylene glycol on glass and for  castor oil on perspex. The coefficients $\beta$ for the best fit are given in table 1.

\begin{figure}[ht]
\begin{center}
\includegraphics[width=14.0cm, angle=0]{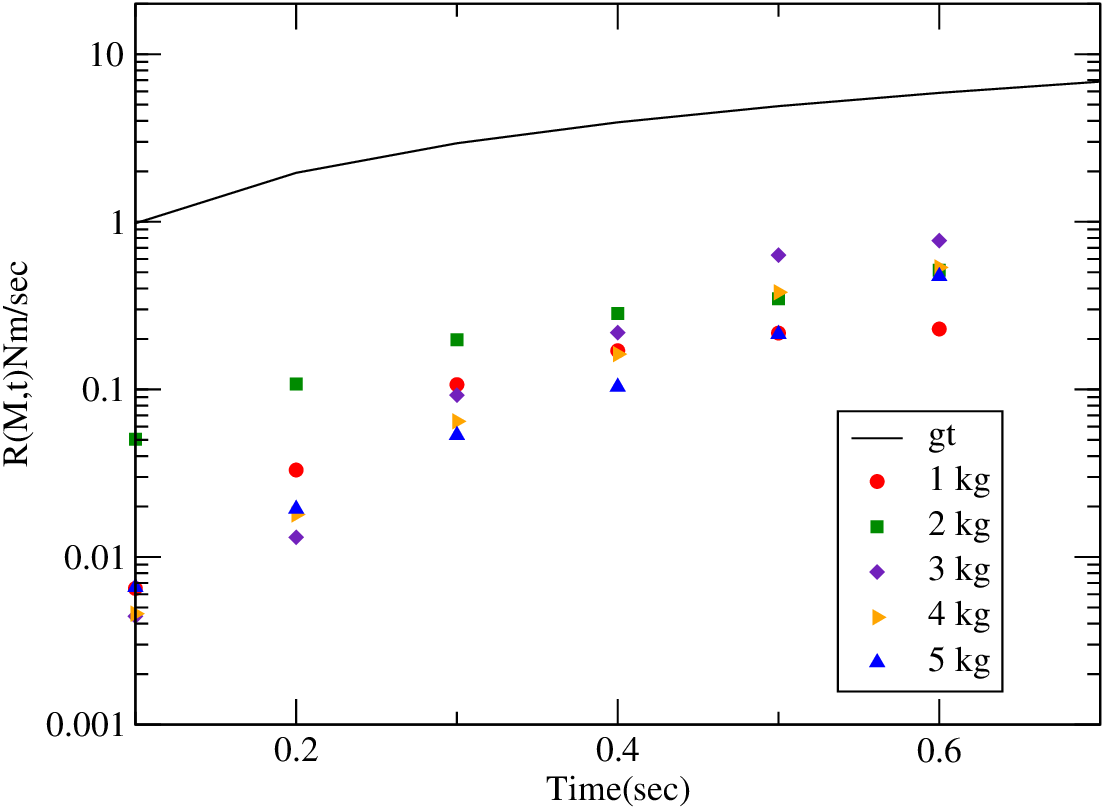}
\end{center} \caption{This figure shows that  gt does not equal R(M,t) for glycol on perspex, for any positive value of $\beta$.} \label{rmt_gp}
\end{figure}

While five of the eight curves can be fit quite well with a finite $\beta$, in the remaining three cases a discrepancy remains. Here the fit  becomes worse for any finite $\beta$. This is shown for glycol on perspex in figure(\ref{rmt_gp}), the other two cases are glycol on glass and linseed oil on perspex.

\section{Discussion}

Out of eight cases studied, we find that the theoretical framework considered here can account for five cases fairly well with one free parameter, the slip coefficient $\beta$. This is not bad considering the complexity of the problem and the simplifications introduced. 

Scrutiny of the table 1, reveals several interesting features. Let us look at $A_0$, the area of contact with only the plate on the fluid drop and no extra weights. we consider this as the reference condition. The mass of the perspex plate is 250 gm and the glass plate is about 700 gms. It is seen from spreading vs. load data (see \cite{apsurf}) that extrapolating to zero load does not change this reference point significantly. We see that for glass $A_0$ ranges from 2-5 cm$^2$, while for perspex we have values less than 1 and the highest is 1.4 cm$^2$. So equilibrium spreading on glass is higher in all cases. However, { \it under load} fluids spread to a much greater extent on perspex, compared to glass. Looking at the percentage (of $A_0$) spreading for the 5 kg. load, we see that for glass it is always less than 50 \%, whatever the fluid, while for perspex the area increases by 100 to 400 \%. So, where the initial spreading is large, the effect of load is less and vice versa.

The highest spreading with load is observed for castor oil on perspex, moreover castor oil shows a creep like behavior and continues to spread for very long times, though the rate slows down. Possibly visco-elastic effects also play important role and should be taken into consideration.

The three cases which cannot be explained by the present theory are all for the low viscosity fluids ethylene glycol and linseed oil. This treatment focussing on the viscous effect works better in case of high viscosity. We have tabulated the dielectric constants along with other data in table 1, because earlier work \cite{epje} showed that the polarizabilties of the fluid and substrate are important in spreading phenomena. But here we find no significant correlation except that the maximum spreading (under load) is for the low dielectric fluid on the low dielectric substrate.

To conclude, spreading of different types of fluids on different solids is a very complex and interesting problem, still open for exploration. Very simple experimental apparatus provides a lot of food for thought in this area. The spreading and flow characteristics are routinely used for classifying soft materials in food technology and chemical engineering, so a deeper understanding will be useful. 

Further experiments and theoretical study taking into account effects of rheology, surface roughness and ambient conditions is required

\begin{table}
\begin{center}
\begin{tabular}{|c|c|c|c|c|c|c|c|c|c|c|}
\hline
solid&fluid&$\epsilon_s$&$\epsilon_f$&$\mu$&S&$\beta$&$A_0$& \% & \% &comment\\
&&&&poise&N/$m^2$&&$cm^2$&spread&spread&\\
&&&&&&&&(1kg)&(5kg)&\\
\hline
\hline
glass&castor&7&4.67&450&3.9&.001&2.05&5&29&good\\
&oil&&&($30^o$C)&&&&&&fit\\
\hline
persp&castor&2.63&4.67&450&3.9&.001&1.42&200&370&good\\
&oil&&&&&&&&&fit\\
\hline
glass&linseed&7&3.35&33&3.3&.028&4.4&11.6&48&good\\
&oil&&&&&&&&&fit\\
\hline
persp&linseed&2.63&3.35&33&3.3&-&0.73&21&135&$\times$\\
&oil&&&&&&&&&\\
\hline
glass&glyce&7&43&1000&6.3&.05&5&11&31&good\\
&rol&&&&&&&&&fit\\
\hline
persp&glyce&2.63&43&1000&63&.0016&1.01&41&90&good\\
&rol&&&&&&&&&fit\\
\hline
glass&glycol&7&41&14&4.8&-&2.22&10&42&$\times$\\
&&&&&&&&&&\\
\hline
persp&glycol&2.63&41&14&4.8&-&0.85&38&235&$\times$\\
&&&&&&&&&&\\
\hline

\end{tabular}
\caption{The table shows properties of the fluids and substrates used, $\mu$ represents the 
viscosity, $\epsilon_s$ and $\epsilon_l$, the solid and liquid dielectric constants, S the surface tension. Results from the experiment are - the initial area of contact $A_0$, the 
percentile spreading of area under loads of 1 and 5 kg., the slip coefficient $\beta$ which
gives a best fit for the spreading master curve.}\label{tabl}
\end{center}
\end{table}

\section{Acknowledgment}
 This work is supported by UGC, Govt. of India. SS is grateful to UGC for award of a research fellowship.

\newpage

\newpage

\newpage

\newpage

\newpage


\begin{thebibliography}{99}
 \bibitem{degen} P. G. de Gennes, Wetting: statics and dynamics, Rev. Mod. Phys. 57(1985) 827-862
 \bibitem{zisman}William A. Zisman, Elaine G. Shaffin, J.Phys.Chem.,1960,64(5),pp 519-524.
 \bibitem{bonn}D. Bonn, J. Eggers, J. Meunier, E. Rolley, Wetting and spreading, to appear in Rev. Mod. Phys. 2009
 \bibitem{engman}J. Engmann, C. Servais, A.S. Burbridge, Squeeze flow theory and applications to rheometry: A review, J. Non-Newt. Fluid Mech. 132(2005)1-27
 \bibitem{asthana}R. Asthana, N. Sobczak, Wettability,spreading and interfacial phenomena in high-temperature coatings, JOM-e 52(2000)
 \bibitem{apsurf}Soma Nag, Suparna dutta, Sujata Tarafdar, Applied Surface Science 256 (2009) 353-355 
 \bibitem{tabor} F.R. Eirich, D. Tabor, Collisions through liquid films, Proc. Camb. Phil Soc. (1948) 566-581
 \bibitem{stefan} J Stefan, Akad.Wiss.Math.Nat.Wien 69(2)(1874)713-735
 \bibitem{epje}S. Sinha, T. Dutta, S. Tarafdar, Eur. Phys. J. E 25(2008) 267-275
 

\end{thebibliography}
\end{document}